\documentclass[prl,superscriptaddress,twocolumn,showpacs,preprintnumbers, 
amsmath,amssymb]{revtex4}

\usepackage{graphicx}
\usepackage{dcolumn}
\usepackage{bm}

\def\bra{\langle}
\def\ket{\rangle}

\begin{document}   
\title{Superconductivity from orbital nematic fluctuations}
\author{Hiroyuki Yamase}
\affiliation{Max-Planck-Institut f\"ur Festk\"orperforschung,
             Heisenbergstrasse 1, D-70569 Stuttgart, Germany}
\affiliation{National Institute for Materials Science, Tsukuba 305-0047, Japan}
\author{Roland Zeyher}
\affiliation{Max-Planck-Institut f\"ur Festk\"orperforschung,
             Heisenbergstrasse 1, D-70569 Stuttgart, Germany}

\date{\today}

\begin{abstract}
Recent experiments suggest that besides of antiferromagnetic fluctuations 
nematic fluctuations may contribute to the occurrence of superconductivity
in iron pnictides. Motivated by this observation we study
superconductivity from nematic fluctuations in a minimal two-band model.
The employed band parameters are appropriate for iron pnictides and 
lead to four pockets for the Fermi line. It is shown that
low-energy, long-wavelength nematic fluctuations within the pockets
give rise to strong-coupling superconductivity whereas the 
large momenta density fluctuations between pockets are rather irrelevant. 
The obtained transition temperatures are similar to those typically found in 
the pnictides and are rather robust against repulsive Coulomb interactions.
The superconducting and nematic states coexist in a large region of the 
phase diagram. 
\end{abstract}


\maketitle
Electronic analogues of nematic liquid crystals attract much interest and are 
observed in a number of condensed matter systems, such as 
quantum spins \cite{penc11}, two-dimensional 
electron gases \cite{lilly99,du99}, cuprates \cite{kivelson03,vojta09}, 
bilayer ruthenates \cite{mackenzie12}, and iron pnictides \cite{fisher11}.  
In a nematic state the orientational symmetry is broken whereas the other 
symmetries are retained. 
Depending on electronic degrees of freedom responsible for nematic order, 
we may distinguish between three kinds of nematicity: 
{\it charge} \cite{kivelson98,yamase00,metzner00}, 
{\it spin} \cite{andreev84,fernandes12}, and {\it orbital} \cite{raghu09,wclee09} 
nematicity. 

In the pnictides a nematic transition is observed just above the 
spin-density-wave (SDW) phase and is
accompanied by a tetragonal-orthorhombic structural phase 
transition \cite{fisher11}. One thus expects both antiferromagnetic
and nematic fluctuations near these transition lines. The nature
of the nematic fluctuations may be either due to spins \cite{fang08,xu08}
or due to orbitals  \cite{krueger09,cclee09,lv09}.
Recent Raman scattering \cite{gallais13} experiments found direct
evidence for low-frequency charge fluctuations  \cite{yamase11,yamase13a}
in this region indicating that these nematic fluctuations involve 
orbital fluctuations even near the SDW phase. Furthermore, magnetic torque 
measurements \cite{kasahara12} showed a nematic transition at higher
temperatures which extends to a region where the SDW is already far away
but where superconductivity still occurs. 

The above experiments suggest an important role of 
nematic fluctuations for the superconductivity in the pnictides. 
A possible non-magnetic mechanism is also suggested by a recent 
discovery that some iron pnictides exhibit superconductivity with critical 
temperature $T_c$ $\sim 40$-$50$ K,  which survives 
very far away from the SDW phase \cite{hanna11,matsuishi12,iimura12}. 

In this paper, we address the question whether orbital nematic fluctuations 
can give rise to high temperature superconductivity by employing 
a minimal two-band model for iron pnictides which 
exhibits four Fermi pockets and a nematic transition. 
We show that interpocket fluctuations which may violate Migdal's theorem
and thus are difficult to calculate
give only 
small contributions to $T_c$. On the other hand intrapocket fluctuations 
can be calculated reliably using the strong-coupling Eliashberg theory, 
which yields $T_c$'s comparable to the observed ones in the pnictides. 
We obtain the superconductivity in a region moderately away from 
the nematic transition and also inside the nematic phase. 

Our approach to superconductivity in pnictides is quite 
different from previous studies \cite{stanescu08,yanagi10a,kontani10,kontani11}, 
although we share their views on the importance of orbital fluctuations. 
In Refs.~\onlinecite{yanagi10a} and \onlinecite{kontani10} 
they considered the weak-coupling limit without quasi-particles 
renormalizations and obtained unrealistic high values for $T_c$  
confined closely to the transition line of orbital order. 
Refs.~\onlinecite{stanescu08} and \onlinecite{kontani11} concluded that  
superconductivity is driven by short-wavelength orbital fluctuations and 
not by long-wavelength ones used by us. 

Our model
Hamiltonian has the form $H = H_0 + H_1$ where the interaction part $H_1$
is given by
\begin{equation}
H_1 = \frac{g}{2} \sum_i n_{i-} n_{i-}.
\label{H1}
\end{equation}
The difference density operator $n_{i-}$ is defined by
$n_{i-} = n_{i1}-n_{i2}$ with the density operator 
$n_{i\alpha} = \sum_\sigma c^\dagger_{i\alpha\sigma} c_{i\alpha\sigma}$. 
$i$ and $\sigma$ are site and spin indices, respectively, and $\alpha=1,2$
is a band index. $g$ is a coupling constant which is considered 
as a parameter in our model. An expression for $H_0$ suitable for pnictides
is \cite{yao09}
\begin{equation}
H_0 = \sum_{{\bf k}, \sigma, \alpha,\beta}\epsilon^{\alpha\beta}_{\bf k}
c^\dagger_{{\bf k}\alpha\sigma} c_{{\bf k}\beta\sigma},
\label{H0}
\end{equation}
with $\epsilon^{11}_{\bf k} = -2t_1 \cos k_x -2t_2 \cos k_y -4t_3 \cos k_x 
\cos k_y$, 
 $\epsilon^{22}_{\bf k} = -2t_2 \cos k_x -2t_1 \cos k_y -4t_3 \cos k_x 
\cos k_y$, $\epsilon^{12}_{\bf k} = -4t_4 \sin k_x \sin k_y$. Reasonable
values for the 
hopping amplitudes are \cite{yao09} 
$t=-t_1,t_2/t=1.5,t_3/t=-1.2,t_4/t=-0.95$, which we will also use in
our calculations; the chemical potential is fixed to be $0.6t$. 
In the following energies are always given in units of $t$.

The above model shows for negative values of $g$ a nematic phase
transition. 
Figure~1 depicts the result of a mean-field calculation of the phase
diagram in the $T$-$g$ plane. Disregarding superconductivity 
the region to the left of the solid line is occupied by the normal state
whereas to the right side of this line a nematic state 
is realized. 
The nematic state is homogenous, i.e., $\bra n_{i-} \ket$ is independent of $i$,  
except for a small region bounded by the line with crosses and the dashed line 
in Fig.~1 where a modulated nematic state is found.

In the following we will study superconductivity from nematic fluctuations. 
We consider the usual Fock diagram for the electronic self-energy 
where the bosonic propagator describes nematic fluctuations. 
Since superconductivity is a Fermi surface effect, we may restrict the  
momenta to the region near the Fermi line of each pocket. In 
our case there are two hole pockets near the $\Gamma$- and M-point and
two electron pockets near the X- and Y-point. We denote them by $i$=1...4, 
respectively, see the inset in Fig. 1. Assuming that the superconducting
order parameter is constant on each individual pocket the Eliashberg
equations for the gap $\Delta_i(i\omega_n)$ and the renormalization
function $Z_i(i\omega_n)$ read as
\begin{equation}
\Delta_i(i\omega_n)Z_i(i\omega_n) = -\pi T\sum_{j,n'}N_j
\frac{\tilde{g}_{ij}(i\omega_n-i\omega_{n'})}{|\omega_{n'}|}
\Delta_j(i\omega_{n'}),
\label{eli1}
\end{equation}
\begin{equation}
Z_i(i\omega_n) = 1-\pi T\sum_{j,n'}N_j\frac{\omega_{n'}}{\omega_n}
\frac{\tilde{g}_{ij}(i\omega_n-i\omega_{n'})}{|\omega_{n'}|}.
\label{eli2}
\end{equation}
$\omega_n$ is a fermionic Matsubara frequency and 
$N_j$ is the density of states of pocket $j$ at the Fermi energy. 
$\tilde{g}_{ij}(i\omega_n-i\omega_{n'})$ is obtained from the microscopic
pairing potential $W_{\alpha\beta}({\bf k},{\bf k'},i\nu_n)$ by
averaging ${\bf k}$ and ${\bf k'}$ independently over the Fermi
lines of pockets $i$ and $j$, respectively, and by summing over
$\alpha$ and $\beta$. $W$ is given by 
\begin{equation}
W_{\alpha\beta}({\bf k},{\bf k'},i\nu_n) = 
V^2_{\alpha\beta}({\bf k},{\bf k'}) g({\bf k}-{\bf k'},i\nu_n), 
\label{W}
\end{equation}
where  $\nu_n$ denotes a bosonic Matsubara frequency. $V$ represents the
form factor
\begin{equation}
V_{\alpha\beta}({\bf k},{\bf k'}) = \sum _{\gamma,\delta}U^\dagger_{\alpha\gamma}({\bf k})
(\tau_3)_{\gamma\delta}U_{\delta\beta}({\bf k'}). 
\label{V}
\end{equation}
$U_{\alpha\beta}({\bf k})$ is a unitary matrix which diagonalizes $H_0$,
$\tau_3$  a Pauli matrix and $g({\bf k}-{\bf k'},i\nu_n)$ is given by
\begin{equation}
g({\bf k}-{\bf k'},i\nu_n) = \frac{g^2\Pi({\bf k}-{\bf k'},i\nu_n)}
{1-g\Pi({\bf k}-{\bf k'},i\nu_n)} +g_0.
\label{g}
\end{equation}
The first term in Eq. (\ref{g}) is the retarded interaction mediated by nematic 
fluctuations, the second one is an instantaneous term accounting for 
repulsive Coulomb interactions. 
$\Pi$ stands for a single bubble of non-interacting nematic particle-hole
excitations.  
\begin{figure}  
\vspace*{0cm}
\includegraphics[angle=0,width=8.0cm]{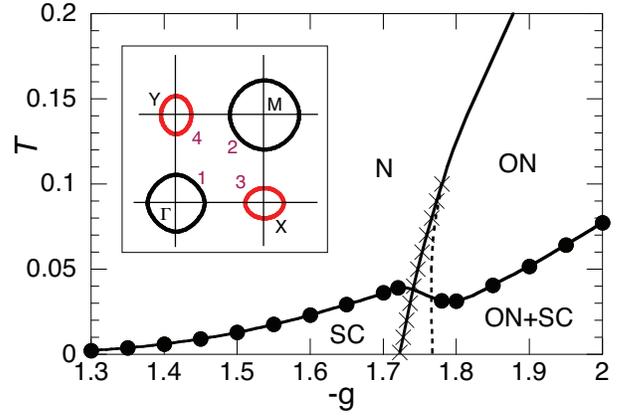} 
\caption{\label{fig:1} 
(Color online) 
Phase diagram in the $T$-$g$ plane for $g_0=0$. 
The thick solid line separates the normal (N) from
the orbital nematic (ON) phase which at low temperatures is modulated  
as indicated by the crosses. The dashed line is the instability line of the 
modulated to the homogeneous nematic state.   
The circles, smoothly joined by a thin solid line, separate
the superconducting (SC) from the normal and nematic states.
The inset shows the Fermi lines of the four pockets in the normal state.
}
\end{figure}
The transition 
temperature $T_c$ to superconductivity is obtained from the condition
that the largest eigenvalue of the matrix 
\begin{equation}
M(i,n;j,n') = 
-\pi T N_j \tilde{g}_{ij}(i\omega_n-i\omega_{n'})/|\omega_{n'}|/Z_i(i\omega_n)
\label{matrix}
\end{equation}
is equal to one.

\begin{figure}[t]  
\vspace*{0cm}
\includegraphics[angle=0,width=7.5cm]{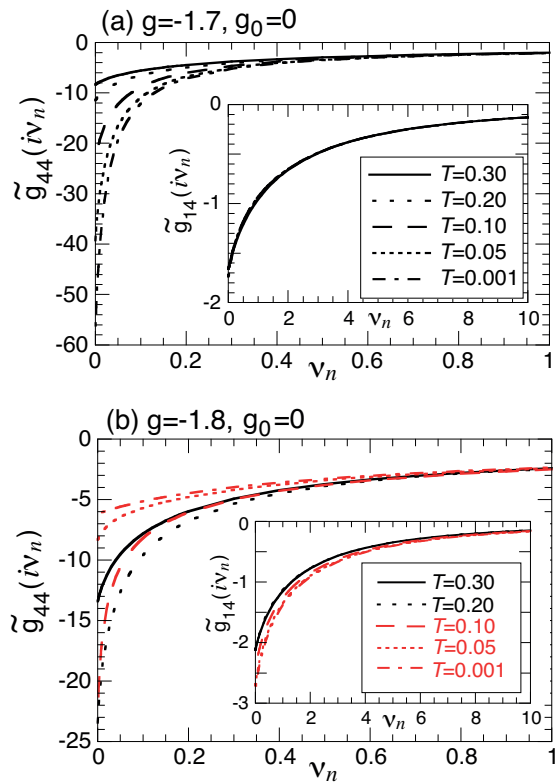} 
\caption{\label{fig:2}
(Color online) 
Retarded pairing interactions $\tilde{g}_{44}$ (main diagrams) and 
$\tilde{g}_{14}$ (insets) as a function of $\nu_n$ for different
temperatures for $g=-1.7$ (a) 
and $g=-1.8$ (b). In the case (b) the nematic state occurs 
below $T=T_n\approx 0.125$. 
}
\end{figure}

The momentum-averaged pairing potentials $\tilde{g}_{ij}(i\nu_n)$
are important ingredients in the calculation of $T_c$. Figure~2 
shows the dependence of $\tilde{g}_{ij}$ on the Matsubara frequency 
$\nu_n$, treated as a continuous variable, for various temperatures. 
The curves for  $\tilde{g}_{44}(i\omega_n)$ 
of Fig.~2 (a) are always in the normal state. With decreasing temperature
the correlation length of nematic fluctuations increases which implies
that  $\tilde{g}_{44}(i\omega_n)$ increases monotonically with decreasing
temperature. At low temperatures $\tilde{g}_{44}(i\nu_n)$ assumes huge
values at $\nu_n=0$ and then decays very fast with increasing frequency
on an energy scale much smaller than $t$. In this frequency range
$\tilde{g}_{44}(i\nu_n)$ varies only slowly with temperature which
indicates that the attractive pairing interaction 
is rather insensitive to temperature and to critical fluctuations.
For $g=-1.8$ [Fig.~2 (b)], the nematic instability occurs at 
the temperature $T_n\approx 0.125$. 
Here  $\tilde{g}_{44}(i\nu_n)$
first increases monotonically with decreasing temperature until 
$T_n$ is reached. Entering the nematic state, 
$\tilde{g}_{44}(i\nu_n)$ is suppressed monotonically due to the development of 
nematic order.  Except for this temperature dependence, 
$\tilde{g}_{44}(i\nu_n)$ is characterized by similar features as in Fig.~2 (a): 
large static values over a wide temperature region, 
a fast decay on an energy scale much smaller than $t$, and a 
weak temperature dependence at high frequencies. 

The insets in Fig.~2 
show the dependence of the non-diagonal interaction $\tilde{g}_{14}$
on $\nu_n$. In this case the momentum average involves large
momenta in the effective interaction and 
the spectral function of the nematic fluctuations extends over a 
large frequency region of several $t$'s. 
As a result the decay of $\tilde{g}_{14}$ with
$\nu_n$ is much slower than in $\tilde{g}_{44}$ and characterized by an 
energy scale of $t$. 
On the other hand the low frequency values
of $\tilde{g}_{14}$ are much smaller than those of $\tilde{g}_{44}$. 
In contrast to the case of  $\tilde{g}_{44}$, the temperature dependence of 
$\tilde{g}_{14}$ is very weak. 
We also
found that the curves for $\tilde{g}_{44}$ and $\tilde{g}_{14}$ in Fig. 2 are
representative for all diagonal $i$=$j$ and non-diagonal $i$$\neq$$j$ interactions, respectively.
The only exception is $\tilde{g}_{34}$ which becomes practically zero due to 
the matrix element $V_{\alpha \beta}({\bf k}, {\bf k'})$ in Eq.~(\ref{W}).  
    
\begin{figure}  
\vspace*{0cm}
\includegraphics[angle=0,width=7.0cm]{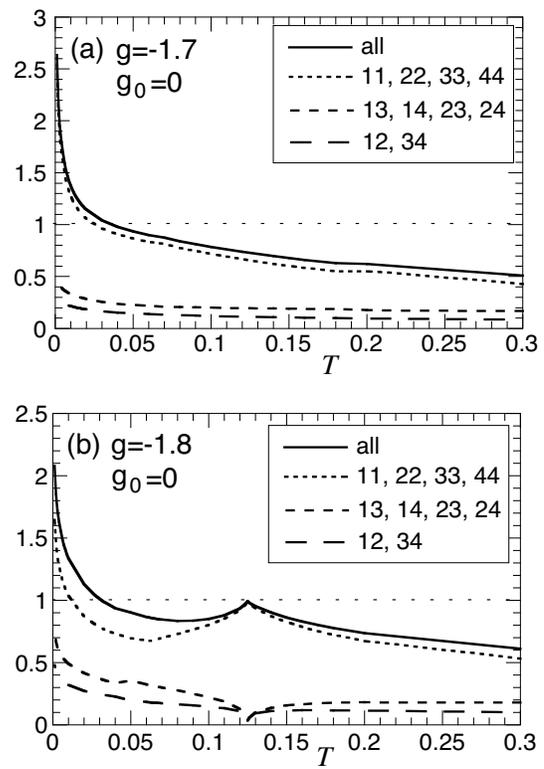} 
\caption{\label{fig:3}
Temperature dependence of the largest eigenvalue using only diagonal 
(dotted line),
non-diagonal (short and long dashed lines) or all pairing 
interactions (solid line).
}
\end{figure}

We have solved Eqs.~(\ref{eli1})-(\ref{eli2}) numerically using up to
2546 Matsubara frequencies to get convergence. Figure~3 (a) shows the largest
eigenvalue of the matrix  $M(i,n;j,n')$
as a function of temperature for $g_0=0$ and $g=-1.7$.
Keeping only the diagonal 11,22,33,44 (non-diagonal 13,14,23,24 or 12,34)
contributions in this matrix, one obtains the dotted (short-dashed or 
long-dashed) curves; the solid line contains all contributions. 
The two curves corresponding to non-diagonal contributions are small 
down to low temperatures and thus do not lead to superconductivity.
The closeness of the solid and dotted lines proves that the total
diagonal contribution is much larger than that of the non-diagonal terms.
$T_c$ is thus determined   
to a large extent by excitations with small momentum transfer. 

To understand the connection between $T_c$ and 
the frequency dependence
of the pairing potentials $\tilde{g}_{ij}$ (Fig.~2), we first note that the
zero frequency value of the pairing potentials with $i$=$j$ does not enter the 
largest eigenvalue when it crosses one. This follows immediately
from Eqs.~(\ref{eli1})-(\ref{eli2}), highlighting the crucial role of $Z_i(i\omega_n)$. 
Thus the strong attractive 
part in the pairing potentials between 0 and $2\pi T_c$ does not contribute 
at all to the instability towards superconductivity. At large temperatures
mainly the flat high-frequency tail of ${\tilde g}_{ij}$ matters 
and already here
the diagonal contribution accounts for most of the largest eigenvalue. 
Decreasing
the temperature the low-frequency behavior of $\tilde{g}_{ij}$ becomes more
and more important because of the substantial increase of  the number of Matsubara 
frequencies contributing to the right-hand side of Eq.~(\ref{eli1}). 
This is the case also for Fig.~2 (b) although the low energy weight is suppressed 
at lower temperature in the nematic phase. 
As a result the huge increase of attraction
towards low frequencies in the diagonal contributions dominates the total
pairing potential, pushes the largest eigenvalue up to one 
and thus determines $T_c$.

Figure~3 (b) shows the largest eigenvalue for $g=-1.8$
where the nematic phase is stable below $T_n \approx 0.125$. The non-diagonal
contributions are again small and do not cross the value one. The 
diagonal and also the total contribution behave similar as in Fig.~3 (a) 
except in a temperature region around $T_n$ where the curves 
approach one from below. This findings follows directly from 
Eqs.~(\ref{eli1})-(\ref{eli2}). At the transition point $T_n$, 
$\tilde{g}_{ii}(i\nu_n=0)$ and $Z_i(i\omega_0)$ tend to 
minus and plus infinity for $i=3$ and 4, respectively; 
$\omega_0$ is the lowest fermionic Matsubara frequency. 
This implies that the largest eigenvalue approaches one from below, 
but never crosses one. 
This could be interpreted in terms of enhanced superconducting fluctuations
near $T_n$. However, the assumptions underlying 
Eqs.~(\ref{eli1})-(\ref{eli2}) could break down
near $T_n$, and non-Fermi liquid features might appear 
similar as in Ref.~\onlinecite{yamase12}.

The pairing potentials ${\tilde g}_{ij}(i\nu_n)$ are for $g_0 = 0$ 
attractive for all 
pocket indices $i,j$ and all Matsubara frequencies. As a consequence
the components of the eigenvector $\Delta_i(i\omega_n)$ belonging to the 
largest eigenvalue have the same sign for all pockets and frequencies.
This means that only the $s_{++}$ symmetry for the order
parameter can be realized. It is interesting that the second largest
eigenvalue lies only a little below the solid line in Fig.~3 in the normal state. 
The corresponding eigenvector has $d$-wave symmetry, i.e., $\Delta_1(i\omega_n)
=\Delta_2(i\omega_n)=0$ and $\Delta_3(i\omega_n)=-\Delta_4(i\omega_n)$ for
all frequencies. To understand this approximate degeneracy, we first take into 
account only the rows and columns related to $\Delta_3$ and $\Delta_4$ 
in our matrix Eq.~(\ref{matrix}). 
Since $\tilde{g}_{34}$ is practically zero, $\Delta_3$ and $\Delta_4$ are 
essentially decoupled and there are two degenerate eigenvectors 
$(\Delta_3, \Delta_4) \propto (1,1)$ and $(1,-1)$, belonging to $s$- and $d$-wave symmetry. 
Considering now also the rows and columns related to $\Delta_1$ and $\Delta_2$ 
the $d$-wave eigenvector does not change by symmetry whereas the $s$-wave 
eigenvector contains contributions from $\Delta_1$ and $\Delta_2$, 
with amplitudes comparable to $\Delta_3$ and $\Delta_4$, leading to our $s_{++}$ state. 
The two eigenvalues are close to each other whereas the eigenvectors are quite different.

The filled circles in Fig.~1, smoothly joined by a thin solid line,
show $T_c$ as a function of $g$. With increasing $-g$, $T_c$ 
monotonically increases because the nematic transition at $g_n \approx -1.72$
is approached and the nematic fluctuations increase. $T_c$ decreases 
somewhat when passing through the transition to the nematic state. 
Increasing $-g$ further enhances $T_c$ monotonically. Although 
the fluctuations are suppressed deep inside the nematic phase, 
they are sufficiently strong to drive a superconducting instability, leading to 
a coexistence of the superconducting and nematic states at lower temperatures. 
In the nematic phase, the Fermi pocket 3 (4) can disappear 
for $\bra n_{i-}\ket >0$ ($<0$) at low temperatures, which is indeed the case 
for $-g \geq 1.8$. Experimentally, such a reconstruction has not been observed so far. 
However, $T_c$ is not affected much by such a Fermi surface reconstruction. 

$Z_i(i\omega_n)$ assumes in general 
values much larger than one at low frequencies, 
which indicates that the strong-coupling limit of superconductivity applies.
Taking the free value $Z_i(i\omega_n)=1$ in Eq.~(\ref{eli1}) would lead in general 
to unphysically high critical temperatures and thus to incorrect conclusions.  
Approximating the frequency dependence of 
$\tilde{g}_{44}$ by a two-square-well model \cite{carbotte90} the cutoff frequency would be 
about $\omega_c \sim 0.05$. Since on the average $T_c \sim 0.03$
we obtain the ratio $T_c/\omega_c \sim 0.6$ which is no longer
small compared to one and thus again points to the 
presence of the strong-coupling limit. $T_c = 0.03$ corresponds to about 50~K
for the hopping $t=150$ meV.
Since $T_c$ is determined by 
intrapocket nematic fluctuations with an energy scale of $\omega_c \sim 0.05$,
which is much smaller than characteristic electronic energies,
Migdal's theorem should be well satisfied. This justifies
our neglect of vertex corrections in Eqs.~(\ref{eli1})-(\ref{eli2}).
If T$_c$ would be dominated by interpocket fluctuations, for instance by 
$\tilde{g}_{14}$, the typical energy scale would be $\omega_c \sim 1$ 
according to Fig. 2 and Migdal's theorem could be violated. 
    
\begin{figure}  
\vspace*{0cm}
\includegraphics[angle=0,width=7.0cm]{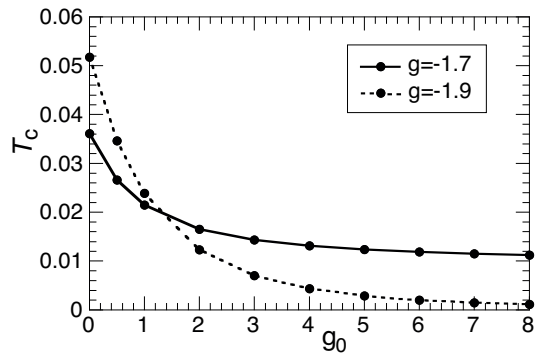} 
\caption{\label{fig:4}
Dependence of $T_c$ on the repulsive, instantaneous interaction $g_0$.
}
\end{figure}

In Fig.~4, we show the dependence of $T_c$ on the instantaneous repulsive 
interaction $g_0$, which will be present in real materials and might 
hinder nematic superconductivity. 
For the case $g=-1.7$ superconductivity evolves from
the normal state and $T_c$ is only moderately suppressed by $g_0$. 
The components of the eigenvector belonging to the largest eigenvalue 
have at low 
frequency the same sign, hence $s_{++}$ symmetry \cite{misc-symmetry},
but change the sign at a finite Matsubara frequency. 
This sign change leads to a partial cancellation of 
the repulsive terms in Eq.~(\ref{eli1}), which allows 
a non-vanishing solution for $\Delta_i(i\omega_n)$ even for a large $g_0$.  
Thus large repulsion terms become ineffective 
by retardation effects in our case, and not by 
non-conventional symmetries for the order parameter, such as $d$-wave. 
For the case of $g=-1.9$ in Fig. 4, superconductivity arises inside 
the nematic phase. Comparing the two panels in Fig. 2 indicates that 
retardation effects are in the nematic 
phase weaker than in the normal phase at low temperatures. 
The cancellation of strong repulsive 
terms in Eq.~(\ref{eli1}) by sign changes of the eigenvector is then
impeded which explains the faster decay of $T_c$ and the vanishing of $T_c$
with increasing $g_0$. 

In conclusion, we have considered a minimal two-band model for iron pnictides 
and found that orbital nematic fluctuations generate strong-coupling 
superconductivity with $T_c$'s comparable to the observed ones in the pnictides. 
In contrast to a spin-fluctuation mechanism \cite{mazin08,kuroki08}, 
which is based on interpocket fluctuations, 
the present approach relies on intrapocket fluctuations 
which could yield a natural explanation for high-$T_c$ with 
a nodeless gap even in the case that only one pocket exists \cite{zhang11a,he13}.  
Moreover, our non-magnetic mechanism 
is supported by the facts that long-wavelength charge fluctuations 
were observed by Raman scattering even close to 
the SDW phase \cite{gallais13}, large $T_c$'s occur in
various pnictides even far away from the SDW 
phase \cite{hanna11,matsuishi12,iimura12}   
and, at least in one case, near a nematic transition line \cite{kasahara12}. 
We also found that it is absolutely necessary to include the renormalization of quasi-particles
due to nematic fluctuations, which was neglected 
in the previous studies \cite{yanagi10a,kontani10,kontani11}. 
As a result $T_c$ is no longer tight closely 
to the nematic transition in agreement with the general phase diagram 
of pnictides \cite{stewart11} 
and superconductivity can occur also inside the nematic phase 
in agreement with the recent experiment \cite{kasahara12}. 

The authors thank A. Eberlein and A. Greco for a critical reading of 
the manuscript and W. Metzner for discussions. 
H.Y. acknowledges support by the Alexander von Humboldt Foundation 
and a Grant-in-Aid for Scientific Research from Monkasho.       
            
\bibliography{main.bib}

\begin{thebibliography}{39}
\bibitem[{mis()}]{misc-symmetry}
\bibinfo{note}{The symmetry changes to $s_{\pm}$ with
  $\Delta_{1,2}(i\omega_0)>0$ and $\Delta_{3,4}(i\omega_0)<0$ if $g_0$ is
  sufficiently large and $g>g_n$.}
\end{thebibliography}



\end{document}